\documentclass[twocolumn,superscriptaddress,prl,aps,preprintnumbers,longbibliography]{revtex4-2}
\usepackage{amsmath,amssymb,bm,graphicx,color,gensymb,bbold,appendix}
\usepackage[colorlinks=true, citecolor={blue!80!black}, urlcolor={blue!50!black}, linkcolor = {blue!80!black}]{hyperref}
\usepackage[utf8x]{inputenc}

\usepackage{xcolor}

\newcommand{\CCoB}{Cs$_2$CoBr$_4$}

\newcommand{\BACOVO}{BaCo$_2$V$_2$O$_8$}
\newcommand{\SRCOVO}{SrCo$_2$V$_2$O$_8$}

\newcommand{\CONBO}{CoNb$_2$O$_6$}
\newcommand{\RbCoCl}{RbCoCl$_3$}

\newcommand{\be}{\begin{equation} }
	\newcommand{\ee}{\end{equation} }
\newcommand{\bea}{\begin{eqnarray} }
	\newcommand{\eea}{\end{eqnarray} }

\usepackage{pdfpages} 
\usepackage{pgffor} 
\makeatletter
\AtBeginDocument{\let\LS@rot\@undefined}
\makeatother
\pdfximage{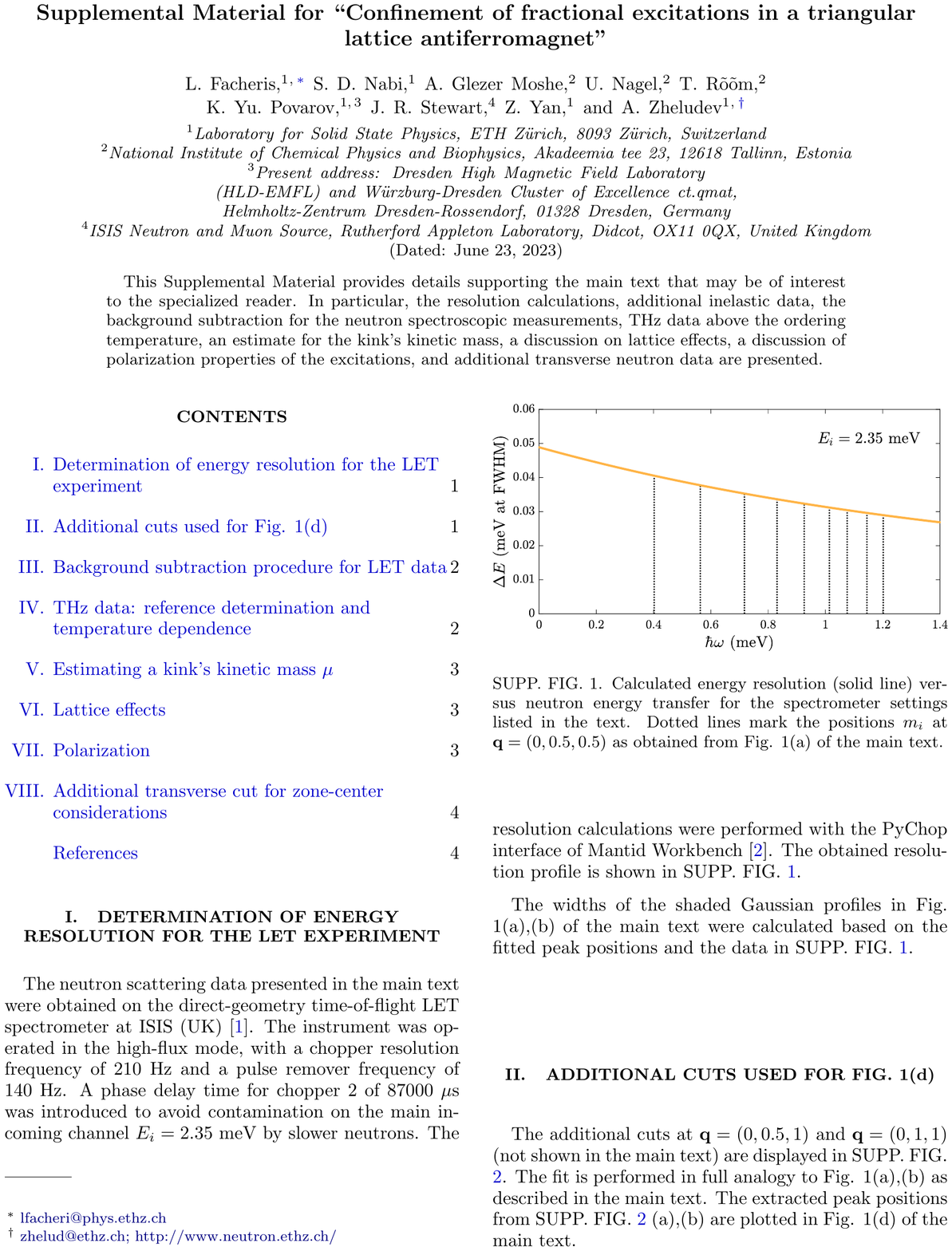}
\def\numbersupplementpages{\the\pdflastximagepages}

\begin{document}

\title{Confinement of fractional excitations in a triangular lattice antiferromagnet}
\author{L.~Facheris}
\email{lfacheri@phys.ethz.ch}
\address{Laboratory for Solid State Physics, ETH Z\"{u}rich, 8093 Z\"{u}rich, Switzerland}
\author{S. D.~Nabi}
\address{Laboratory for Solid State Physics, ETH Z\"{u}rich, 8093 Z\"{u}rich, Switzerland}
\author{A.~Glezer~Moshe}
\author{U.~Nagel}
\author{T.~R{\~o}{\~o}m}
\address{National Institute of Chemical Physics and Biophysics, Akadeemia tee 23, 12618 Tallinn, Estonia}
\author{K.~Yu.~Povarov}
\address{Laboratory for Solid State Physics, ETH Z\"{u}rich, 8093 Z\"{u}rich, Switzerland}
\address{Present address: Dresden High Magnetic Field Laboratory (HLD-EMFL) and W\"urzburg-Dresden Cluster of Excellence ct.qmat, Helmholtz-Zentrum Dresden-Rossendorf, 01328 Dresden, Germany}
\author{J. R. Stewart}
\address{ISIS Neutron and Muon Source, Rutherford Appleton Laboratory, Didcot, OX11 0QX, United Kingdom}
\author{Z. Yan}
\address{Laboratory for Solid State Physics, ETH Z\"{u}rich, 8093 Z\"{u}rich, Switzerland}
\author{A.~Zheludev}
\email{zhelud@ethz.ch; http://www.neutron.ethz.ch/}
\address{Laboratory for Solid State Physics, ETH Z\"{u}rich, 8093 Z\"{u}rich, Switzerland}

\begin{abstract}
High-resolution neutron and THz spectroscopies are used to study the magnetic excitation spectrum of \CCoB, a distorted-triangular-lattice antiferromagnet with nearly XY-type anisotropy. What was previously thought of as a broad excitation continuum [Phys. Rev. Lett. \textbf{129}, 087201 (2022)] is shown to be a series of dispersive bound states reminiscent of ``Zeeman ladders'' in quasi-one-dimensional Ising systems. At wave vectors where inter-chain interactions cancel at the Mean Field level, they can indeed be interpreted as bound finite-width kinks in individual chains. Elsewhere in the Brillouin zone their true two-dimensional structure and propagation are revealed.
\end{abstract}

\date{\today}
\maketitle

In conventional magnetic insulators the dynamic response is typically dominated by coherent single-particle $S=1$ excitations, aka magnons or spin waves. In many low-dimensional and highly frustrated quantum spin systems elementary excitations carry fractional quantum numbers, be they spinons in Heisenberg spin chains \cite{FaddeevTachtajan_PLA_1981_spinwaves,Muller_PRB_1981_ansatz,Stone_PRL_2003_CupzNcontinua,Giamarchi_2004_1Dbook}, Majorana fermions in the now-famous Kitaev model \cite{Kitaev_AnnPhys_2006_Kitaev,WinterTsirlin_JPCM_2017_KitaevReview,TakagiTakayama_NatRevPhys_2019_KitaevReview}, or pseudo-charge defects in spin ice-like Coulomb phases~\cite{Henley2010}. The physical excitation spectrum, such as that measured by neutron spectroscopy, is then dominated by broad multi-particle continua~\cite{MourigalEnderle_NPhys_2013_4spinon,Schmidiger_PRL_2013_DimpyLET,DaiZhang_PRX_2021_NaYbSespinonFS,Tennant_JPSJ_2019_QMneutronReview,Plumb2019}. In addition to the continuum, fractional excitations may also form bound states due to attractive interactions between them. A spectacular new phenomenon emerges when interactions are {\em confining}, i.e. do not fall off with distance, much like strong forces that bind quarks in hadrons~\cite{Wilczek_AnnRevNuclPart_1989_QCDreview}. This produces a hierarchical series of bound states inside the resulting potential well. An example is the sequence of domain wall (kink) bound states in quasi-one-dimensional (quasi-1D) Ising spin chains~\cite{McCoyWu_PRD_1978_IsingKinks,Shiba_PTP_1980_ZeemanLadder,Rutkevich_JStPhys_2008_ConfinedKinks}. The confining potential for this model is linear and results from 3D couplings, which generate an effective field acting on individual chains~\cite{Shiba_PTP_1980_ZeemanLadder}. The binding energies are, in supreme mathematical elegance, spaced according to the negative zeros of the Airy function \cite{McCoyWu_PRD_1978_IsingKinks,Rutkevich_JStPhys_2008_ConfinedKinks}.
The best-known experimental examples of such ``Zeeman ladder'' spectra are the quasi-1D Ising ferromagnet \CONBO~\cite{Coldea_Zeemanladder} and antiferromagnet (AF) \BACOVO~ \cite{GrenierPetit_PRL_2015_BACOVOzeemanladders}, as well as the isostructural compound \SRCOVO~\cite{BeraLake_PRB_2017_SrCoVOZeemanLaddersNeutrons}, where as many as 8 consecutive bound states are observed. Shorter sequences have been found in another prototypical Ising spin chain material, \RbCoCl~\cite{MenaRuegg_RbCoCl3_PRL}. 

Bound states are always easier to form in 1D, where even an arbitrary-shallow potential well provides at least one robust localized state for a propagating particle. Fractional excitations can also form bound states in higher dimensional spin systems \cite{Matan2022,Ghioldi2022}. Their structure and propagation are much more complex than that of simple paired domain walls in spin chains. Unfortunately, no Zeeman ladder analogues, i.e. no hierarchical binding sequences of fractional excitations have been found in higher dimensions to date. In the present work we report the observation of such a phenomenon in a {\em quasi-2D} distorted-triangular-lattice AF. That the quintessentially 1D physics of bound kinks survives in 2D is remarkable. We argue that it is ``rescued'' at certain special wave vectors by the intrinsic frustration in triangular lattice geometry. Elsewhere in the Brillouin zone the bound states are no longer restricted to single chains and are to be viewed as 2D objects propagating in the entire triangular plane.

\begin{figure}
	\includegraphics[width=\columnwidth]{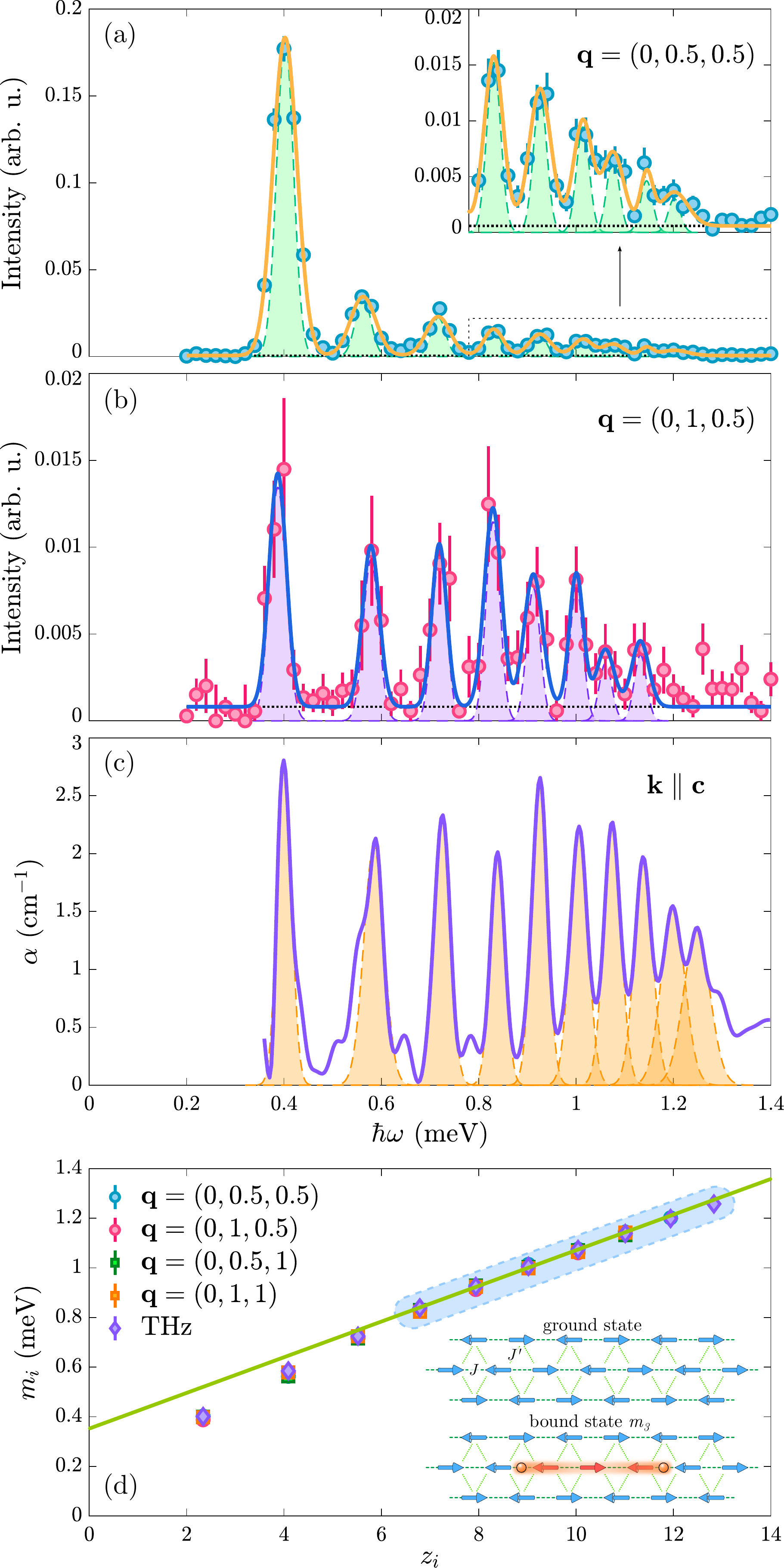}
	\caption{(a)-(b) Neutron scattering intensity (solid symbols) measured at $ T =40$~mK versus energy transfer at the 1D AF zone-centers $\mathbf{q}= (0, 0.5, 0.5)$ and $\mathbf{q} = (0, 1, 0.5)$, respectively. The data are integrated fully along $h$ direction and in $\pm 0.025$~r.l.u. and $\pm 0.25$~ r.l.u. along $k$ and $l$, respectively. Solid lines are fits to a series of Gaussian peaks. Dashed Gaussians represent the calculated experimental energy resolution. Black dotted lines indicate the fitted flat background. (c) Measured THz absorption (solid line) versus absorbed photon energy for light propagating along the $\bm{c}$ axis at $0.2$~K. Dashed areas highlight the Gaussian components of the fit to extract peak positions. (d) Measured excitation energy plotted versus the value of negative roots of the Airy function. The solid line is a linear fit as described in the text. The blue area highlights the points used for the fit. Inset: cartoons of the magnetic ground state and a representative $m=3$ 2-kink bound state.}\label{ladder}
\end{figure} 

The material in question, \CCoB~(space group P$nma$, $a = 10.19$, $b = 7.73$, $c = 13.51$ \AA ), is a very interesting $J-J'$ model distorted-triangular-lattice AF \cite{PovarovFacheris_PRR_2020_CCoBplateaux,FacherisPovarov_PRL_2022_SDWUUD}. Despite a prominent triangular motif in its structure, it demonstrates certain 1D features such as a field-induced incommensurate spin density wave with Tomonaga-Luttinger spin liquid type dynamics and a propagation vector controlled by a one-dimensional nesting in the spinon Fermi sea. Its true 2D nature is manifest in the presence of a robust $m=1/3$ magnetization plateau, typical of a triangular AF. The model magnetic Hamiltonian is described in detail in Refs. \cite{PovarovFacheris_PRR_2020_CCoBplateaux,FacherisPovarov_PRL_2022_SDWUUD}. The key structural features are chains of Co$^{2+}$ ions that run along the crystallographic $\bm{b}$ axis of the orthorhombic lattice (see Fig. 1 in Ref. \cite{PovarovFacheris_PRR_2020_CCoBplateaux}). The chains are coupled in the $(bc)$ plane in a zigzag fashion to form a distorted triangular network (inset of Fig.~\ref{ladder}(d)). Easy-plane single-ion anisotropy ensures that the low-energy physics of the spin-$3/2$ Co$^{2+}$ ions can be described in terms of effective $S=1/2$ pseudo-spins. The components of the effective exchange coupling constants are subject to restrictions imposed by the pseudo-spin projection. A simplistic spin-wave analysis of previous inelastic neutron data provided a rough estimate for the nearest-neighbor in-chain AF exchange tensor components: $J^{XX}\sim J$, $J^{YY}\sim 1.1J$, $J^{ZZ}\sim 0.25J$, $J = 0.8$~meV \cite{FacherisPovarov_PRL_2022_SDWUUD}. Here $Y$ is chosen along the $\bm{b}$ crystallographic direction, and $X$ and $Z$ alternate between adjacent chains, where anisotropy planes are almost orthogonal. Note that this is practically a {\em planar exchange anisotropy}, with only a tiny in-plane Ising component to account for the $\Delta\sim 0.4$~meV spectral gap found in this system. The frustrated {\em inter-chain coupling $J'$ is significant}, of the order of $0.45 J$, and is of predominantly Ising ($YY$) character. Inter-plane interactions $J''$ are not frustrated. The material orders magnetically in a collinear stripe-type structure, with an ordering wavevector $(0, 1/2,1/2)$ (see inset in Fig.~\ref{ladder}(d)). The N\'{e}el temperature $T_\text{N}=1.3$~K allows us to estimate $J''$. If this were the only coupling between chains with no additional frustration due to $J'$, we could expect $ k_\mathrm{B}T_\mathrm{N}\sim 2\Delta/\ln(\Delta/J'')$~\cite{CarrTsvelik_PRL_2003_XXZchainRPA}. The actual value of $J''$ must be larger than thus obtained, as the in-plane frustration interferes with the emerging magnetic structure. A certain upper estimate is given by the mean field picture where $k_\mathrm{B}T_\mathrm{N}\sim 2J''S(S+1)$. This leads us to conclude that $3\cdot10^{-4}~\mathrm{meV}\lesssim J''\lesssim 0.075$~meV $\ll J$, confirming the quasi-2D character of the material.

Our previous inelastic neutron scattering experiments indicated that the excitation spectrum in zero applied field is a gapped continuum of states, with intensity concentrated on its lower bound, and a strong dispersion along the chain axis \cite{FacherisPovarov_PRL_2022_SDWUUD}. The central finding of the present work is that this ``continuum'' is actually a sequence of at least 9 sharp bound states that previously could not be observed due to poor experimental energy resolution. New neutron data were collected at the LET time-of-flight spectrometer at ISIS (UK), using $2.35$~meV incident energy neutrons in repetition-rate-multiplication mode \cite{LET_bewley}. We used the same $1.16$~g single crystal as in \cite{FacherisPovarov_PRL_2022_SDWUUD} mounted on a $^3$He-$^4$He dilution refrigerator. All measurements were performed at a base temperature of $40$~mK. In the experiment the sample was rotated $180^\circ$ around the $\bm{a}$ axis in steps of $1^\circ$. The spectra were measured for $\sim10$-minute counting time at each sample position.

We first focus on the 1D AF zone-centers ($\mathbf{q}\bm{b}=0,\pi$), where inter-chain interactions within the triangular planes {\em cancel out at the Mean Field-RPA level}, and where spin wave theory predicts no transverse dispersion or intensity modulation of excitations. Fig.~\ref{ladder}(a),(b) show constant-$\mathbf{q}$ cuts through the data at wave vectors $\mathbf{q}=(0,0.5,0.5)$ and $\mathbf{q}=(0,1,0.5)$, respectively. A sequence of sharp peaks is clearly apparent in both cases. A fit to the data using empirical Gaussian profiles yields an accurate measure of the peak positions and shows that their widths are essentially resolution-limited. In Fig.~\ref{ladder}(a),(b) this is emphasized by the shaded Gaussians representing the computed experimental resolution \cite{SM}\nocite{MantidPaper}\nocite{Horace_Ewings}\nocite{SWTHzBook2018}\nocite{FurrerMesotBook}.

Corroborative evidence is also obtained by THz spectroscopy. The experiment was performed with a Martin-Puplett-type interferometer and a $^3$He-$^4$He dilution refrigerator with base temperature of $150$~mK using a $^3$He-cooled Si bolometer at $0.3$\,K. The sample was a circular plate approximately $1$~mm thick in $\bm{c}$ direction and $4$~mm in diameter. THz radiation propagating along the crystal $\bm{c}$ axis was unpolarized and the apodized instrumental resolution was $0.025$~meV. The THz absorption spectrum is shown in Fig.~\ref{ladder}(c). It is calculated as a difference of spectra measured at $0.2$~K and $2$~K, i.e. in the magnetically ordered phase and above $T_\text{N}$. The THz spectrum appears to have some features absent in the neutron spectrum \footnote{The extra features are superficially similar to field-induced satellites observed in ~\cite{AmelinTHz2020}, but appear in zero field in our case. They are more likely related to different polarization channels, as discussed below.}, but all peaks found in the latter are also present here.  The positions of these peaks were determined in Gaussian fits (shaded peaks) in a narrow range $\pm 0.025$~meV near each peak value. Comparing data measured at low temperature to that collected at $2$~K~\cite{SM} reveals that the series of peaks is endemic to the ordered phase and disappears above $T_\text{N}$.

The spacing between the excitation peaks present in both measurements corresponds to confinement in an approximately linear one-dimensional potential. To demonstrate this, we plot the excitation energies deduced from neutron spectra at several wave vectors, as well as the positions of corresponding THz peaks, versus the negative roots $z_i$ of the Airy function in Fig.~\ref{ladder}(d). For a precise linear attractive potential $\lambda |x|$ between the dispersive particles \footnote{See discussion of lattice effects in the Supplemental Material.}, near the minimum $\epsilon(k)=m_0+\hbar^2 k^2 /2\mu$ we expect the excitation energies to be~\cite{Coldea_Zeemanladder,Rutkevich_JStPhys_2008_ConfinedKinks}

\begin{equation}
m_i=2m_0+(\hbar\lambda)^{2/3}\mu^{-1/3} z_i\,\,\,\,\text{with}\,\,\,\,i=1,2,\dots.
\end{equation}

In the actual data, the linear dependence is apparent for all but the first few points. As will be addressed in more detail below, this slight deviation indicates that the confining force increases somewhat at short distances. From a linear fit to the higher-energy peaks we can immediately extract the slope $0.072(3)$~meV and the energy of a single particle $m_0=0.18(1)$~meV (half-intercept). Using the single-particle kinetic mass $\hbar^2 /\mu=0.39$~meV$\times b^2$~\cite{SM}, we estimate the confining force constant $\lambda=0.031(2)$~meV/$b$ \footnote{This force of $\sim 6$~fN corresponds to the gravity pull between two average humans at a separation of $8$~km.}.

\begin{figure}
	\includegraphics[width=\columnwidth]{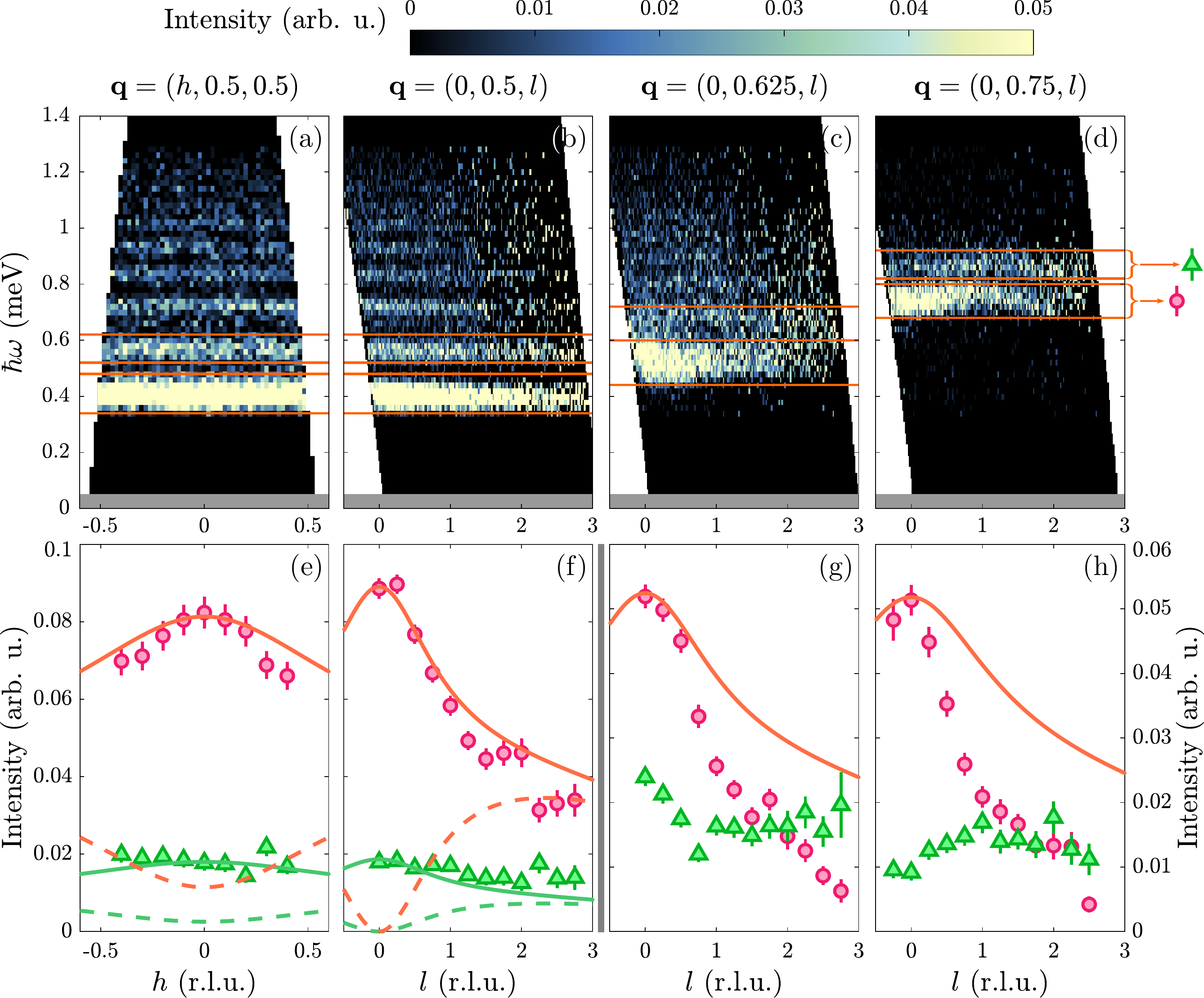}
	\caption{(a)-(d) False color plot of neutron scattering intensity measured at $ T = 40$~mK plotted versus
energy transfer and momentum transfer transverse to the crystallographic $\bm{b}$-axis. Gray areas mask regions of elastic-incoherent scattering. Background subtraction has been performed as described in \cite{SM}. The orange regions represent energy-integration windows used to extract the cuts in panels below. (e)-(h) Intensity-momentum cuts (solid symbols) for the first two modes. The orange and green lines show the product of the magnetic form-factor-squared for Co$^{2+}$ and calculated neutron polarization factor for excitations polarized perpendicular (solid) and parallel (dashed) to the ordered moment direction, respectively. The overall scale factor of the lines is arbitrary, but always consistent between the two polarizations and consistent between panels (e) and (f).}\label{spectra}
\end{figure}

The next point that we make is that the observed bound states at the 1D AF zone-center are essentially 1D objects. This is concluded by analyzing the neutron spectra shown in Figs.~\ref{spectra}(a),(b) and Supp. Fig.~5(a) in~\cite{SM}. The bound states {\em do not propagate in either transverse direction} and thus have an essentially flat dispersion. Moreover, their intensity shows a monotonic falloff and no obvious transverse modulation, as plotted for the first two modes in Figs.~\ref{spectra}(e),(f), and Supp. Fig.~5(b)~\cite{SM}. This implies that these excitations do not involve cross-chain correlations and are {\em confined to a single chain}.

For the first mode, the measured transverse wave vector dependence is precisely accounted for (orange solid lines in Figs.~\ref{spectra}(e),(f), and Supp. Fig.~5(b)) by the combined effects of i) the magnetic form factor of Co$^{2+}$ and ii) a neutron polarization factor for spin components perpendicular to the chain axis (to the direction of ordered moments in the ground state). Longitudinal excitations would become stronger for scattering vectors deviating from the chain axis (orange dashed line). Already for the second mode the identification of polarization is ambiguous (green lines). Unlike in Ising chains, in \CCoB~ the conservation of the chain-axis projection of angular momentum is not expected, meaning that excitations cannot be unambiguously classified as longitudinal nor transverse.

The 1D nature of zone-center excitations prompts a simple interpretation. Similarly to the situation in \CONBO~and \BACOVO, the observed modes are bound states of two kinks (domain walls) in individual chains. Naively, ignoring the kinetic energy of the kinks as in the purely Ising case, such an excitation can be visualized as in the cartoon in the inset of Fig.~\ref{ladder}(d). The energy $m_0$ is to be associated with that of a single domain wall. As a consistency check, we can compare that to the computed energy of a domain wall in a classical spin chain. Using $J^{YY}/J^{XX}\sim 1.1$ as estimated for \CCoB, with a trivial numerical classical-energy minimization procedure we get $m_0\sim 0.9 JS^2 = 0.18$~meV, in excellent agreement with the measured value.

Geometric frustration ensures that at the magnetic zone-center these strings of flipped spins within a single chain incur no energy cost due to interactions with adjacent chains within the triangular lattice. Moreover, any transverse dispersion is Suppressed. At the same time, the interaction energy due to unfrustrated inter-layer coupling is proportional to the string length, resulting in confinement. In this simplistic picture, the confining force is $\lambda=2J''S^2/b$. This yields an inter-layer coupling constant $J''=0.062(4)$~meV, inside the possible range deduced from $T_\mathrm{N}$. 

The apparent difference between our data and those for the Ising-chain AF \BACOVO~\cite{GrenierPetit_PRL_2015_BACOVOzeemanladders} is the presence of {\em two} Zeeman ladder series in the latter material. The two series correspond to transverse and longitudinal excitations, the first transverse mode being a spin flip, i. e. a conventional magnon. They are offset in energy by $\lambda b$ in our notation~\cite{SM}. For \CCoB~this amounts to as little as $0.03$~meV. The two series can not be resolved within the resolution and counting statistics of our neutron experiment but may account for the extra shoulder peaks seen in the THz data in Fig.~\ref{ladder}(c). A similar estimate for \BACOVO~yields a larger offset of about $0.34$~meV that is easily resolved experimentally \cite{GrenierPetit_PRL_2015_BACOVOzeemanladders}.

The deviation from linear-potential behavior at low energies is also readily explained. Since the material is almost planar, the domain walls are not confined to a single bond as in the ideal Ising case, but have a characteristic size $l$ \footnote{We define the domain wall width in a spin-$S$ chain as the distance over which the $z$-axis spin component changes from $S/2$ to $-S/2$ near its center.}. We can estimate that quantity in a classical spin chain using the above-mentioned anisotropy parameters: $l\sim 2b$. The energy of the first few bound states is thus modified due to a physical overlap of the two bounding domain walls. Once the kinks are separated by a distance of more than $\sim l$, this interaction becomes negligible and the confinement potential becomes linear.

In summary, at the 1D AF zone-center the simplistic kink model from 1D Ising physics consistently explains the observed excitation energies. The first mode is transversely polarized and shows an intensity patterns very consistent with that of a magnon in spin wave theory~\cite{FacherisPovarov_PRL_2022_SDWUUD}, being strongly suppressed at the 1D structural zone-centers (integer $k$). The intensities of higher modes are not that easily interpreted: they may have a mixed polarization and are less suppressed at integer $k$ values (Fig.~\ref{ladder}(a),(b)).

\begin{figure}
	\includegraphics[width=\columnwidth]{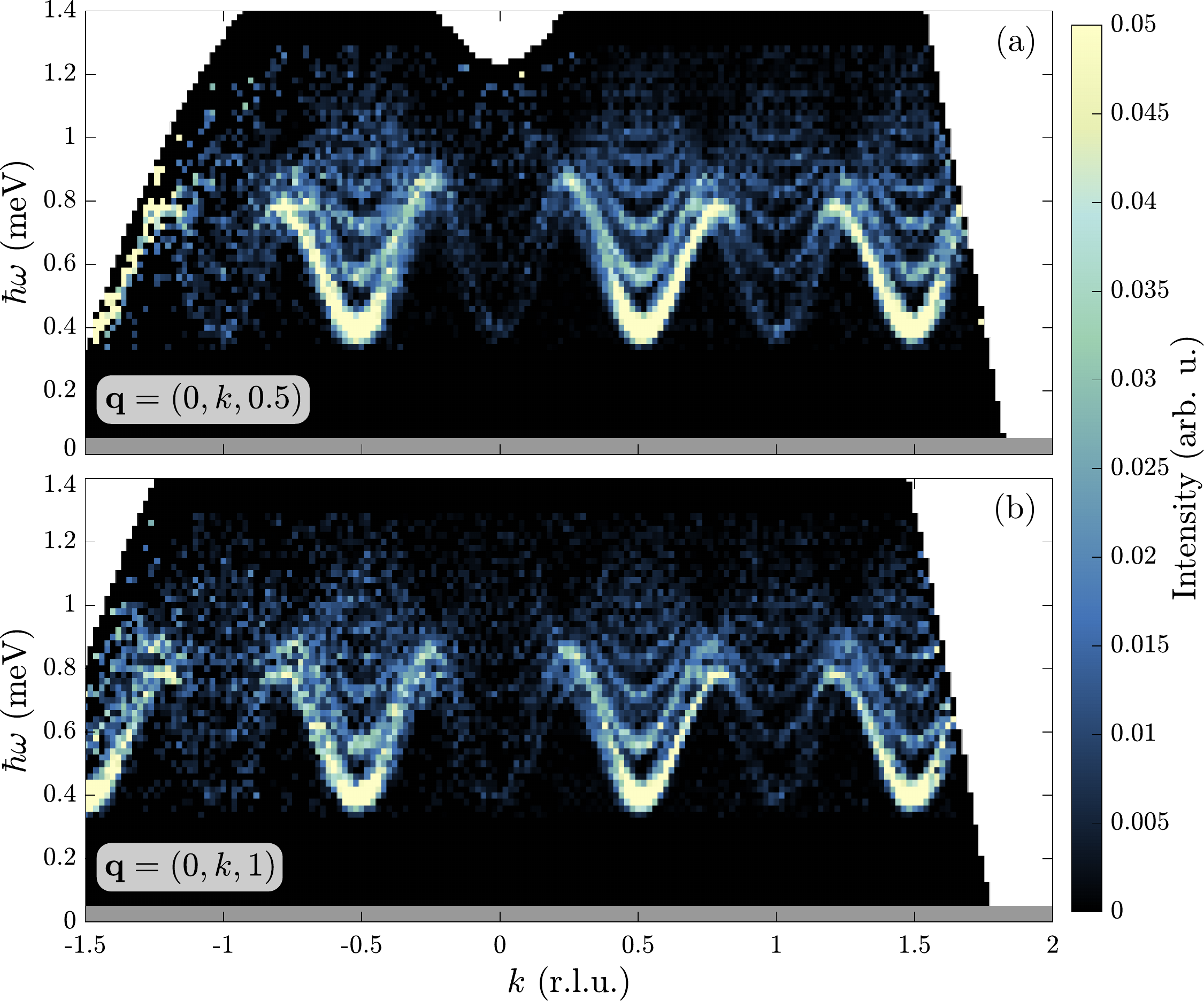}
	\caption{(a)-(b) False color plot of neutron scattering intensity measured at $T = 40$~mK plotted versus energy transfer and momentum transfer along $\mathbf{q} = (0, k, 0.5)$ and $\mathbf{q} = (0, k, 1)$ respectively. The data were fully integrated along $h$, and in the range $\pm 0.25$~r.l.u. along $l$ around the central value. The gray areas mask regions where the incoherent scattering dominates the signal. Background subtraction has been performed as described in \cite{SM}.}\label{spectrum}
\end{figure}

Away from the 1D AF zone-centers, the excitations are considerably more complex. This is very clear in the longitudinal dispersion of the bound states shown in Fig.~\ref{spectrum}(a),(b). Other than at $\mathbf{q}\bm{b}=0,\pi$ ($k = 0, 1/2$) the $m_1$ mode splits into two branches, each with an asymmetric dispersion relation. In fact, the $m_1$ state at $\mathbf{q}\bm{b}=\pi$ seems to be continuously connected to the $m_2$ excitations at $\mathbf{q}\bm{b}=2\pi$ ($k = 1$) and vice versa. Fitting the dispersion of the strongest low-energy mode in the vicinity of $\mathbf{q}\bm{b} = \pi$ to a Lorentz-invariant relativistic form
\be
{(\hbar\omega_\mathbf{q})}^2=\hbar^2 \Delta \,(\mathbf{q}\bm{b})^2\,/\mu +\Delta^2,
\ee
yields the value of kinetic mass quoted above.

A look at the intensities reveals that other than at the special wave vectors, the bound states can no longer be seen as strings in a single chain, but are ``dressed'' with correlations extending to several neighboring chains in the triangular plane. This conclusion is reached from Fig.~\ref{spectra}(c),(d), that show a transverse cut of the spectrum at $\mathbf{q}\bm{b}=5\pi/4$ and $\mathbf{q}\bm{b}=3\pi/2$, respectively. As plotted in Fig.~\ref{spectra}(g),(h), the measured intensity of the first two modes now shows a much steeper transverse wave vector dependence than computed from just the polarization and form factors (solid lines). The second mode even seems to show signs of intensity oscillations, with a possible mixing/hybridization and resulting intensity transfer with the first mode.

Our data reveal that away from the special wave vectors the bound states also {\em propagate} in two dimensions, albeit with a small bandwidth. Indeed, in Fig.~\ref{spectra}(d) one can see that at $\mathbf{q}\bm{b}=3\pi/2$ the bound states develop a non-zero dispersion along the $\bm{c}^*$ direction, in contrast to what is seen at $\mathbf{q}\bm{b}=0,\pi$. Although the bandwidth of transverse dispersion, $0.08$~meV, is at the limit of our experimental resolution, qualitatively one can say that $\mathbf{q}\bm{c}=0,4\pi$ are dispersion minima for the $m_1$ mode, while the maximum is at $\mathbf{q}\bm{c}=2\pi$. That periodicity is consistent with having two chains per unit cell along the $\bm{c}$-axis direction in the crystal structure.

Overall, the differences between our results and spectra of Ising spin chains \cite{Coldea_Zeemanladder,GrenierPetit_PRL_2015_BACOVOzeemanladders} are striking. In the latter, all bound states, including the first one, are much less dispersive than the lower edge of the entire spectrum, which approximately corresponds to the lower edge of the two-kink continuum in the absence of long-range order. As a result, each bound state persists only in a restricted area in the Brillouin zone. In contrast, in \CCoB~a few of the lower-energy bound states are highly dispersive and span across the entire zone.

In summary, we demonstrate that ``Zeeman ladders'' of confined fractional excitations can exist in a {\em bona fide} quasi-2D system. These states are inherently related to those in 1D models, as revealed at special wave vectors where 2D interactions are canceled by geometric frustration. However, elsewhere in reciprocal space their true 2D character is manifest. Once again, the distorted triangular lattice model provides a link between 1D and 2D quantum magnetism.
\begin{acknowledgements} This work was Supported by a MINT grant of the Swiss National Science Foundation. We acknowledge Support by the Estonian Research Council grants PRG736 and MOBJD1103, and by European Regional Development Fund Project No. TK134. Experiments at the ISIS Neutron and Muon Source were Supported by beam time 
allocation RB2210048 from the Science and Technology Facilities Council \cite{ISIS_data}.
\end{acknowledgements}

\bibliography{Zeeman_Cs2CoBr4_18-05-2023}



\section*{Supplementary material}

# Supplemental Material for "Confinement of fractional excitations in a triangular lattice antiferromagnet"

L. Facheris,[1, *] S. D. Nabi,[1] A. Glezer Moshe,[2] U. Nagel,[2] T. Rõõm,[2]
K. Yu. Povarov,[1, 3] J. R. Stewart,[4] Z. Yan,[1] and A. Zheludev[1, †]

[1]*Laboratory for Solid State Physics, ETH Zürich, 8093 Zürich, Switzerland*
[2]*National Institute of Chemical Physics and Biophysics, Akadeemia tee 23, 12618 Tallinn, Estonia*
[3]*Present address: Dresden High Magnetic Field Laboratory
(HLD-EMFL) and Würzburg-Dresden Cluster of Excellence ct.qmat,
Helmholtz-Zentrum Dresden-Rossendorf, 01328 Dresden, Germany*
[4]*ISIS Neutron and Muon Source, Rutherford Appleton Laboratory, Didcot, OX11 0QX, United Kingdom*
(Dated: June 23, 2023)

This Supplemental Material provides details supporting the main text that may be of interest to the specialized reader. In particular, the resolution calculations, additional inelastic data, the background subtraction for the neutron spectroscopic measurements, THz data above the ordering temperature, an estimate for the kink's kinetic mass, a discussion on lattice effects, a discussion of polarization properties of the excitations, and additional transverse neutron data are presented.

## CONTENTS

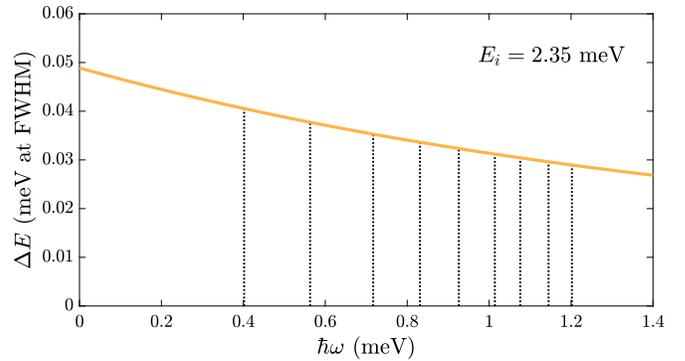

SUPP. FIG. 1. Calculated energy resolution (solid line) versus neutron energy transfer for the spectrometer settings listed in the text. Dotted lines mark the positions $m_i$ at $\mathbf{q} = (0, 0.5, 0.5)$ as obtained from Fig. 1(a) of the main text.

## I. DETERMINATION OF ENERGY RESOLUTION FOR THE LET EXPERIMENT

The neutron scattering data presented in the main text were obtained on the direct-geometry time-of-flight LET spectrometer at ISIS (UK) [1]. The instrument was operated in the high-flux mode, with a chopper resolution frequency of 210 Hz and a pulse remover frequency of 140 Hz. A phase delay time for chopper 2 of 87000 $\mu$s was introduced to avoid contamination on the main incoming channel $E_i = 2.35$ meV by slower neutrons. The resolution calculations were performed with the PyChop interface of Mantid Workbench [2]. The obtained resolution profile is shown in SUPP. FIG. 1.

The widths of the shaded Gaussian profiles in Fig. 1(a),(b) of the main text were calculated based on the fitted peak positions and the data in SUPP. FIG. 1.

## II. ADDITIONAL CUTS USED FOR FIG. 1(d)

The additional cuts at $\mathbf{q} = (0, 0.5, 1)$ and $\mathbf{q} = (0, 1, 1)$ (not shown in the main text) are displayed in SUPP. FIG. 2. The fit is performed in full analogy to Fig. 1(a),(b) as described in the main text. The extracted peak positions from SUPP. FIG. 2 (a),(b) are plotted in Fig. 1(d) of the main text.

---

* lfacheri@phys.ethz.ch
† zhelud@ethz.ch; http://www.neutron.ethz.ch/

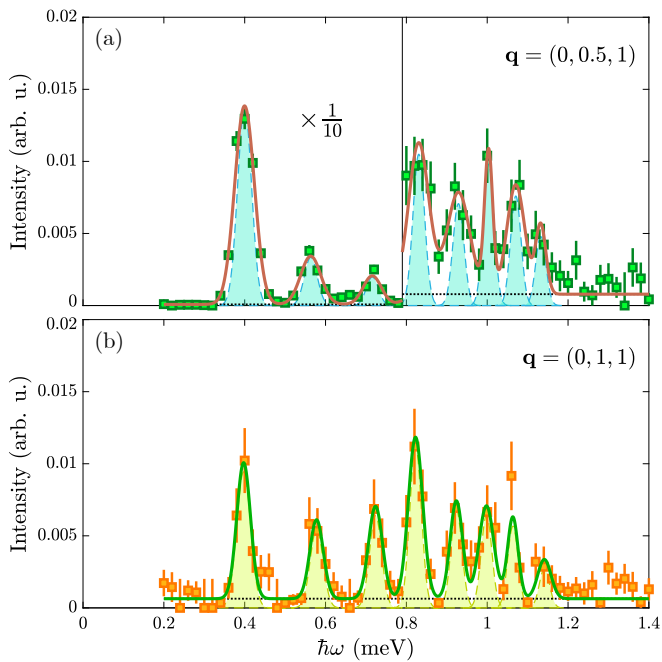

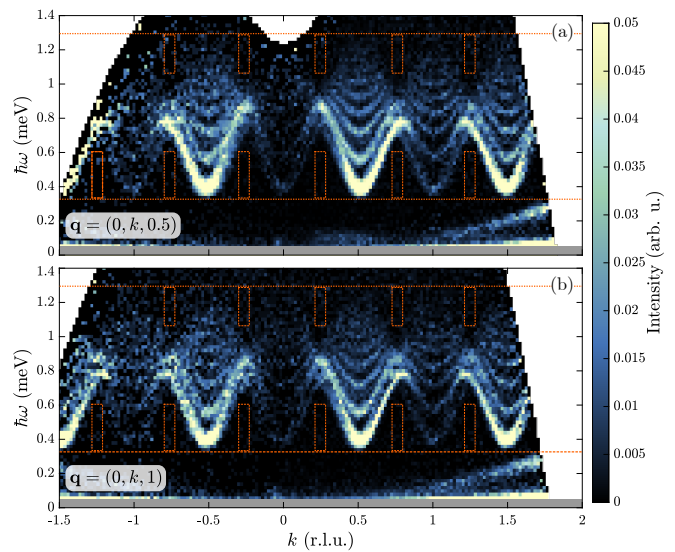

SUPP. FIG. 2. (a)-(b) Neutron scattering intensity (solid symbols) measured at $T = 40$ mK versus energy transfer at $\mathbf{q} = (0, 0.5, 1)$ and $\mathbf{q} = (0, 1, 1)$, respectively. The data are integrated fully along $h$ direction and in $\pm 0.025$ r.l.u. and $\pm 0.25$ r.l.u. along $k$ and $l$, respectively. Solid lines are fits to a series of Gaussian peaks. Dashed Gaussians represent the calculated experimental energy resolution. Black dotted lines indicate the fitted flat background.

SUPP. FIG. 3. (a)-(b) False color plot of raw neutron scattering intensity measured at $T = 40$ mK plotted versus energy transfer and momentum transfer along $\mathbf{q} = (0, k, 0.5)$ and $\mathbf{q} = (0, k, 1)$ respectively. The data were fully integrated along $h$, and in the range $\pm 0.25$ r.l.u. along $l$ around the central value. The gray areas mask regions where the incoherent scattering dominates the signal. Orange dashed lines and boxes delimit the edges of the background dataset, as described in the text.

## III. BACKGROUND SUBTRACTION PROCEDURE FOR LET DATA

The inelastic neutron scattering data presented in Fig. 2 and Fig. 3 of the main text are background-subtracted. Although the dataset was rather clean, a background subtraction similar to that in [3] was nonetheless performed. In this section the model adopted to describe the background is outlined. The analysis was performed using the Horace software package [4].

SUPP. FIG. 3 shows raw data corresponding to Fig. 3 of the main text. Strong sharp lines at the edges of the dataset below 0.4 meV are known spurious originating from scattering from the sample environment employed. The total background was modeled assuming no magnetic scattering below the gap and above the top of the spectrum. Thus, the background dataset is identical to original data for $\hbar\omega \leq 0.34$ meV and $\hbar\omega \geq 1.28$ meV (see dashed horizontal lines in SUPP. FIG. 3 for the background regions projected on these particular cuts). In the intermediate energy region, momentum-dependent boxes were constructed as shown in SUPP. FIG. 3 and numerically interpolated over the total explored $(\mathbf{q}, \hbar\omega)$-space. The so-obtained background was then point-to-point subtracted from the original data.

## IV. THZ DATA: REFERENCE DETERMINATION AND TEMPERATURE DEPENDENCE

The measurement of THz absorption spectra always relies on reference profiles collected using the same setup [5]. The usual way to obtain such a reference consists in measuring the intensity transmitted through an hole. However, this may not always provide the best transmission spectrum, because of the difference in interference and diffraction effects between the sample and the hole [5]. This is especially true if the absorption lines of interest are weak.

An alternative way consists in performing differential measurements using the same sample under different external conditions (e.g. high temperature, high magnetic field) [5]. As stated in the main text, the spectra shown in Fig. 1(c) use a measurement at 2 K (above $T_\mathrm{N}$) as a reference. This latter approach of course relies on the assumption that above the ordering temperature the magnetic signal is absent or at least very broad and featureless. In order to verify this, we performed additional measurements in strong magnetic field. The idea is that in magnetic field substantially exceeding the pseudospin saturation field $\mu_0 H_c = 6$ T the magnon spectrum is gapped and the low-energy THz signal contains no magnetic contribution. Given the known value of the gyromagnetic ratio $g \sim 2.4$ [6], we can be fairly certain that in a field of $\mu_0 H = 12$ T applied along the **c** axis this will be the case

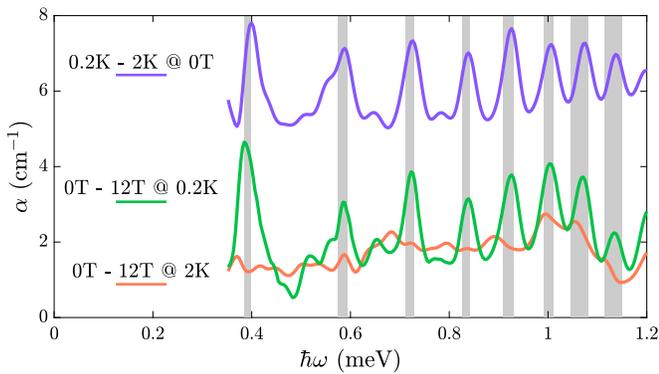

SUPP. FIG. 4. Measured THz absorption versus photon energy for light propagating along the **c** axis in zero-field at $T = 0.2$ K (green solid line) and 2 K (orange solid line). These two lines were referenced to spectra measured in 12 T applied field at the respective temperature. The purple line is the same temperature-normalized data as in Fig. 1(c) for comparison, shifted upward for visibility by 5 cm$^{-1}$. Grey vertical bars indicate peak positions with relative uncertainties from the neutron data set.

at least up to 1.5 meV. In SUPP. FIG. 4 we show both $T = 0.2$ K and $T = 2$ K spectra in zero field referenced to 12 T at the respective temperatures. For comparison, we also show the temperature-differential data set of Fig. 1(c) of the main text. Regardless of which reference data set is used, high field or high temperature, the "Zeeman ladder" is clearly visible at $T = 0.2$ K. At $T = 2$ K it is no longer observable, which validates our approach.

## V. ESTIMATING A KINK'S KINETIC MASS $\mu$

Near its minimum at a 1D wave vector $k_0 = \frac{\pi}{b}$, the dispersion relation for a single kink can be approximated as

$$\epsilon_k = m_0 + \frac{\hbar^2}{2\mu}(k - k_0)^2. \tag{S.1}$$

The parameter $\mu$ is the kinetic "mass" of this quasiparticle. We can access it from the experimentally measured spectrum of two-kink excitations. For a two-kink state, energy-momentum conservation dictates

$$\hbar\omega_q^{(2-\text{kink})} = \epsilon_k + \epsilon_{q-k} = 2m_0 + \frac{\hbar^2}{2\mu}\left[(k - k_0)^2 + (q - k + k_0)^2\right] \tag{S.2}$$

Minimizing (S.2) with respect to the "hidden" quasi-momentum $k$, we find that the lower boundary of the two-particle continuum lies at $k = q$. Thus, the lowest magnon-like dispersion is given by:

$$\hbar\omega_q = 2m_0 + \frac{\hbar^2}{2\mu}\left[(q - k_0)^2 + k_0^2\right] \tag{S.3}$$

Near the minimum wave vector $q_0 = k_0 \to \pi/b$, we find that the curvature of the parabola-like dispersion is actually the same for a single kink and the lowest bound state.

## VI. LATTICE EFFECTS

Eq. (1) in the main text is the result for a particle confined in a linear potential well *in free space*. It brings with itself a characteristic length scale $l$ that roughly corresponds to the physical size of the first (smallest) bound state. From simple dimensional analysis $l = [\hbar^2/(\mu\lambda)]^{1/3}$. In a spin system we have a second length scale, namely the lattice spacing $b$. Strictly speaking, Eq. (1) is only applicable if $l \gg b$. However, even if this criterion is not satisfied, the higher the bound state, the greater its physical size, and the less significant any lattice effects will be.

Another way to reach the same conclusion is by comparing the characteristic energy spacing between low-index bound states in the continuum model, which is of the order of $(\hbar\lambda)^{2/3}\mu^{-1/3}$, with the potential drop across one lattice site, namely $\lambda b$. For light particles ("kinetic limit") the former term dominates and the continuum model can be applied without reservation. In the opposite "static" limit the motion of kinks can be disregarded. One can take the cartoon in the inset of Fig. 1(d) literally, as it will correspond to an (almost) exact eigenstate of the system. In this case, the index of the bound state is equal to the integer length of the string connecting the kinks. The energy levels become uniformly spaced, with a step $\lambda b$.

In our particular case of $Cs_2CoBr_4$, based on the estimated $\lambda$ and $\mu$, we have $l/b \sim 2$. This is clearly pushing the self-consistency criterion, and may be contributing to the deviation from the linear behavior seen in Fig. 1(d). On the other hand, as mentioned above, higher-index bound states will be progressively less affected by the discretization introduced by the lattice, validating our approach.

Additional re-assurance can be drawn from a comparison with the simpler and well-characterized Ising-chain system $BaCo_2V_2O_8$ [7]. First we fit our Eq. (2) to the bottom of the dispersion measured in that material to obtain $\hbar^2/\mu = 1.10$ meV$\times c^2$, where $c$ is the corresponding chain axis lattice parameter. From the reported coefficient $\alpha \sim 0.42$ meV between $m_i$ and $z_i$ [7], we then obtain $\lambda = 0.3$ meV/$c$. This gives $l/c \sim 1.5$, smaller than in our case. And yet, in $BaCo_2V_2O_8$ a linear relation between bound state energies and Airy function roots holds remarkably well in the entire range.

## VII. POLARIZATION

Due to spin, the problem of confined kinks in an AF Ising chain is subtly different from that of a single particle

in a potential well. In an AF with XXZ symmetry, the net $z$-axis spin projection is conserved. Kinks carry $S_z = \pm 1/2$, and thus can form three Zeeman ladder sequences of bound states with $S_z = \pm 1$ or $S_z = 0$. The two former are degenerate by virtue of XXZ symmetry, but the third may be distinct. In a neutron scattering experiment they correspond to transverse and longitudinal excitations. As explained in Ref. [7, 8], in the static limit longitudinal and transverse excitations correspond to kinks separated by even and odd number of sites, respectively.

In the static limit, the confinement of kinks in longitudinal excitations is mapped to confinement of spinless particles on a lattice with spacing $2b$. For odd-length transverse-polarized bound states, we can again map the problem to particles on a lattice with spacing $2b$, but there is an additional constant contribution to the energy equal to $-\lambda b$. We end up with two Zeeman ladders that are offset by that energy relative to one another. While this argument applies only in the static limit, we can use it to at least roughly estimate the splitting between longitudinal and transverse modes in the general case.

In $BaCo_2V_2O_8$ both longitudinal and transverse Zeeman ladders are observed and are separated by $0.23 \pm 0.05$ meV [7]. This splitting is large enough to effortlessly resolve in a cold-neutron spectroscopic experiment [9]. It is indeed quite close to our estimate $\lambda c = 0.3$ meV for $BaCo_2V_2O_8$.

For $Cs_2CoBr_4$ $\lambda b \sim 0.03$ meV. This is less that the intrinsic FWHM energy resolution of the TOF neutron spectrometer. Thus, the two series can not be resolved, also because of the relatively high noise level, and appear as a single sequence. On the other hand, both the resolution and signal-to-noise ratios of the THz data are considerably better. The shoulders seen for the first three peaks in the THz data may indeed be related to the second series. Once again we emphasize that due to lack of axial symmetry, the excitations can not be unambiguously classified as "longitudinal" and "transverse" in our material.

## VIII. ADDITIONAL TRANSVERSE CUT FOR ZONE-CENTER CONSIDERATIONS

Additional data supporting the claim on the peculiar behavior at zone center are presented in SUPP. FIG. 5. This shows the spectrum measured versus $\mathbf{q} = (0, 1.5, l)$, integrated and binned in the same manner as Fig. 2(b). Again, the dispersion of the bound states perpendicular to the chain at $k = 1.5$ is essentially flat. The first state's intensity, shown in SUPP. FIG. 5(b) as pink solid symbols, shows a monotonic falloff which is completely captured by the combined effect of polarization factor for transverse excitations and polarization factor (solid orange line). The overall scale factor for this line is the same as in Fig. 2(e),(f) of the main text.

---

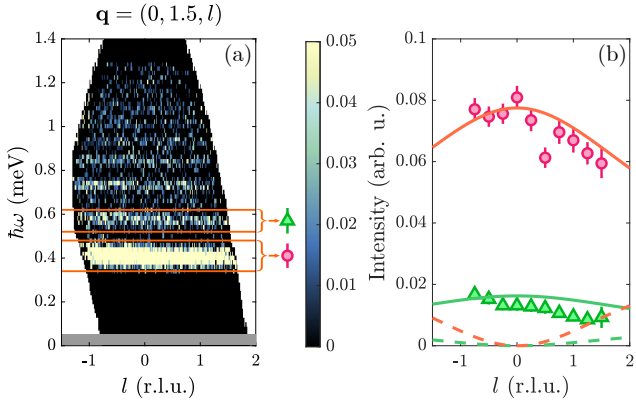

SUPP. FIG. 5. False color plot of neutron scattering intensity measured at $T = 40$ mK plotted versus energy transfer and momentum transfer transverse to the crystallographic **b**-axis. Gray areas mask regions of elastic-incoherent scattering. Background subtraction has been performed as described above. The orange regions represent energy-integration windows used to extract the cut on the right. (b) Intensity-momentum cut (solid symbols) for the first two modes. The orange and green lines show the product of the magnetic form-factor-squared for $Co^{2+}$ and calculated neutron polarization factor for excitations polarized perpendicular (solid) and parallel (dashed) to the ordered moment direction, respectively. The overall scale factor of the lines is consistent between the two polarizations and consistent with panels (e) and (f) of Fig. 2 of the main text.


    
\end{document}